\documentclass[aps,prb,reprint,showpacs,superscriptaddress]{revtex4-1}

\usepackage{amsmath, amssymb, bm, braket, dsfont, times, mathtools,graphicx,color}
\usepackage{tikz}
\usetikzlibrary{arrows}
\usepackage[breaklinks,colorlinks,allcolors=blue]{hyperref}

\usepackage[normalem]{ulem}

\begin{document}

\title{Quantum dynamics of the square-lattice Heisenberg model}
\author{Ruben Verresen}
\affiliation{Department of Physics, T42, Technische Universit\"at M\"unchen, 85748 Garching, Germany}
\affiliation{Max-Planck-Institute for the Physics of Complex Systems, 01187 Dresden, Germany}
\author{Frank Pollmann}
\affiliation{Department of Physics, T42, Technische Universit\"at M\"unchen, 85748 Garching, Germany}
\date{\today}
\author{Roderich Moessner}
\affiliation{Max-Planck-Institute for the Physics of Complex Systems, 01187 Dresden, Germany}
\begin{abstract}
Despite nearly a century of study of the $S=1/2$ Heisenberg model on the square lattice, there is still disagreement on the nature of its high-energy excitations. By tuning toward the Heisenberg model from the exactly soluble Ising limit, we find that the strongly attractive magnon interactions of the latter naturally account for a number of spectral features of the Heisenberg model. This claim is backed up both numerically and analytically. Using the density matrix renormalization group method, we obtain the dynamical structure factor for a cylindrical geometry, allowing us to continuously connect both limits. Remarkably, a semi-quantitative description of certain observed features arises already at the lowest non-trivial order in perturbation theory around the Ising limit. Moreover, our analysis uncovers that high-energy magnons are localized on a single sublattice, which is related to the entanglement properties of the ground state.
\end{abstract}

\maketitle

At its ripe age of 90 years\cite{Heisenberg28}, the square lattice antiferromagnetic Heisenberg model has had its dynamical properties studied intensely. Spin wave theory was in large part developed to investigate this model's low-energy properties\cite{Hulthen36,Anderson52,Harris71}. Its anomalous terms lead to a `relativistic' (linear) low-energy dispersion, related to coupling the two sublattices of the ground state's spontaneous N\'eel ordering\cite{Keffer53}. While evading an exact treatment, its dynamics has been studied numerically via quantum Monte Carlo simulations\cite{Sandvik01,Shao17} and exact diagonalization\cite{Luescher09}; also several high-order perturbative expansions have been devised, such as around the Ising limit\cite{Singh89,Gelfand96,Singh95,Zheng05} or via continuous unitary transformations (CUT)\cite{Powalski15,Powalski17}. In addition, a number of ad-hoc approaches have been motivated by a range of different physical pictures. Moreover, this model is related to experimental systems, not least to the parent states of the cuprates\cite{Anderson87,Rice87,Chakravarty88,Ho01}, the study of which has led to much of the modern theory of quantum magnetism.

It may thus seem all the more surprising that there is still no consensus on the appropriate physical picture for certain regions in momentum space. Proposals include strongly-interacting magnons\cite{Powalski15,Powalski17}, deconfined spinons\cite{Hsu90,Ho01,Syljuaasen02,DallaPiazza15,Ferrari18}, all the way to a connection to deconfined quantum criticality \cite{Shao17}. The disagreement is not limited to the underlying physical mechanism, but also pertains to quantitative aspects of spectral properties. One uniting factor, at least, is a need to go beyond a perturbatively-dressed single-magnon picture.

Our intention is not to propose yet another scenario. Rather, we adopt the perspective that away from the unassailable hydrodynamic limit---accounting for the low-energy Goldstone modes\cite{Keffer53,Nielsen76}---other features which have caught the attention of the community may not even be uniquely described by one picture as opposed to another. Instead, we seek to provide a simple account of salient features of the intermediate and high-energy part of the spectrum.

Perhaps the most controversial region concerns magnons with momenta $|k_x|+|k_y| = \pi$. Spin wave theory predicts that these are dispersionless, in disagreement with both experiment\cite{Ronnow01,Christensen07,DallaPiazza15} and theoretical methods. What is instead observed, is a local mininum at $\bm k = (\pi,0)$---commonly referred as the \emph{roton mode}, in analogy with the quasi-particle dispersion in liquid Helium\cite{Landau41}. Moreover, spin wave theory is unable to account for the high spectral weight in the continuum just above this mode.

In this paper, we advance along two complementary tracks. First, we determine the dynamical structure factor using a method based on the density matrix renormalization group (DMRG)\cite{White92,White93,Schollwoeck11,Stoudenmire13,Zaletel15,Gohlke17}, with systematic errors distinct from those of previous approaches. This novel method has at least two welcome features: it confirms that the phenomenology of the roton mode is indeed beyond the dressed single-magnon picture, and it uncovers a hitherto-unrecognized property of magnons with $|k_x|+|k_y| = \pi$, which we refer to as \emph{sublattice-localization}. We also clarify how the latter is related to the entanglement of the ground state.

Second, we use existing data from an Ising expansion developed by Singh and Gelfand\cite{Singh89,Gelfand96,Singh95}---and pushed further by Zheng, Oitmaa and Hamer\cite{Zheng05}---to point out that some known results at the isotropic point are already semi-quantitatively accounted for by the lowest non-trivial order. Moreover, our numerical method allows us to study such an XXZ model (with dominant easy-axis anisotropy) without any perturbative approximation.

The main message of our paper is that aspects of the attractive magnon interactions, i.e. the physics beyond spin wave theory, arise naturally from domain-wall-counting in the Ising limit, sometimes connecting all the way to the isotropic Heisenberg point. In particular, the numerics shows this at a phenomenological level, but a simple perturbative calculation also sheds light on, e.g., the small-yet-nonzero magnitude and shape of the roton mode's dispersion. Moreover, even the aforementioned phenomenon of sublattice-localization can be accounted for within a low-order perturbative picture. In addition, we provide a quantitative analysis of the roton mode.

The remainder of this paper is structured as follows. In section~\ref{sec:model} we give a brief overview of the model's salient features, relating them to previous literature whenever possible. The spectral functions obtained using DMRG are shown in section~\ref{sec:spectral}: first for the Heisenberg model, which is then connected to the Ising limit. Section~\ref{sec:perturbative} supplements this by showing how various features, such as the roton minimum or sublattice-localization, naturally arise within a low-order perturbative picture. The apparently hitherto-unexplored phenomenon of sublattice-localization is studied numerically in section~\ref{sec:entanglement}, emphasizing its link to entanglement (or absence thereof). Section~\ref{sec:roton} contains a quantitative analysis of the roton mode with comparison to results from the literature.

\section{Square lattice Heisenberg model} \label{sec:model}

We study the spin-$\frac{1}{2}$ antiferromagnetic Heisenberg model (AFH) on the square lattice, allowing for easy-axis anisotropy:
\begin{equation}
H = J \sum_{\langle\bm n, \bm m \rangle} \Big( S^z_{\bm n} S^z_{\bm m} + \lambda \left[S ^x_{\bm n} S^x_{\bm m} + S^y_{\bm n} S^y_{\bm m} \right] \Big) \label{eq:model}
\end{equation}
where $J>0$. We are principally interested in the isotropic point $\lambda=1$, where the N\'eel order of the ground state spontaneously breaks the $SU(2)$ symmetry down to a $U(1)$ group generated by $S^z_\textrm{tot} = \sum_{\bm n} S^z_{\bm n} $ (where we define the ordering direction to be along the spin $z$-axis). As we will argue, it is also useful to consider $0 \leq \lambda < 1$, where the model is in a gapped Ising phase which spontaneously breaks the $\mathbb Z_2$ symmetry $R^x_\pi = \prod_{\bm n} \exp{\left(-i \pi S^x_{\bm n} \right)}$.

\subsection{Dynamical structure factor and quantum numbers}

Spectral functions give direct insight into the properties of excitations. In this work we focus on the dynamical structure factor, which is experimentally accessible through, for example, inelastic neutron scattering. It can be expressed in terms of the dynamical correlation functions $C^{\gamma\gamma}(\bm r,t)  = \langle \sigma^\gamma_{\bm r}(t) \sigma^\gamma_{\bm 0}(0) \rangle = 4\langle S^\gamma_{\bm r}(t) S^\gamma_{\bm 0}(0) \rangle $:
\begin{equation}
\mathcal S^{\gamma \gamma}(\boldsymbol k,\omega) = \frac{1}{2\pi} \sum_{\mathbf r} \int_{-\infty}^\infty  e^{i(\omega t-\mathbf k \cdot \mathbf r)} C^{\gamma\gamma}(\mathbf r, t) \; \mathrm dt, \label{eq:spectral}
\end{equation}
which is normalized as $\int \mathcal S^{\gamma \gamma}(\boldsymbol k,\omega) \; \mathrm d \mathbf k \mathrm d \omega = (2\pi)^d$. We focus on the transverse spectral function:
\begin{equation}
\mathcal S^t(\bm k,\omega) = \mathcal S^{xx}(\bm k,\omega) + \mathcal  S^{yy}(\bm k,\omega) . \label{eq:transverse}
\end{equation}

This object gives direct insight into the excitations above the ground state. If $\gamma =x,y$, one can show\footnote{To argue that \unexpanded{$\langle 0| \tilde S^\gamma_{-\bm k} |\alpha \rangle \langle \alpha | S^\gamma_{\bm 0} |0\rangle = \langle \tilde S^\gamma_{-\bm k} |\alpha \rangle \langle \alpha | \tilde S^\gamma_{\bm k} |0\rangle$}, one can use that \unexpanded{$\tilde S_{\bm k}^\gamma$} has a well-defined quantum number with respect to the \emph{modified} translation operator \unexpanded{$R^\gamma_\pi T_{x,y}$}.} that
\begin{equation}
\mathcal S^{\gamma \gamma} (\bm k, \omega) = \sum_\alpha \delta(\omega - (\omega_\alpha-\omega_0)) \; |\langle \alpha | \tilde S^\gamma_{\bm k} |0\rangle|^2 \label{eq:Lehmann}
\end{equation}
where $\tilde S^\gamma_{\bm k} = \sum_{\bm r} e^{i\bm{k\cdot r}} S_{\bm r}^\gamma$. It is natural to choose a basis $|\alpha\rangle = |\bm k , S^z_\textrm{tot}, \beta\rangle$, where $\bm k$ is the momentum with respect to the translation symmetry $T_{1,\pm 1}$ of the \emph{two-site unit cell}. Eq.~\eqref{eq:Lehmann} tells us that the spectral function gives information about the existence of energy eigenstates with momentum $\bm k$ and and $S^z_\textrm{tot} = \pm 1$.

Note that when labeling states, $\bm k$ lives in the reduced (magnetic) Brillouin zone, $|k_x| + |k_y| \leq \pi$, but the spectral function itself is periodic only with respect to the original (lattice) Brillouin zone, $-\pi \leq k_x, k_y \leq \pi$ (taking the lattice constant to be unity).

\subsection{Spin wave theory}

In terms of the above quantum numbers, spin wave theory predicts two bands\footnote{One could instead work in a local frame where there is only one band, but this obscures some of the physics. In particular, then $S^z_\textrm{tot}$ is no longer a good quantum number.}. These exactly coincide and are distinguished by $S^z_\textrm{tot} = \pm 1$. The dispersion relation to order\footnote{To wit, the Holstein-Primakoff transformation is $S^+ = \sqrt{2S}\sqrt{1-\frac{a^\dagger a}{2S}} a$. The dominant term of the square root $\sqrt{1-\frac{a^\dagger a}{2S}}$ is of order $1/S^0$, which leads to LSWT.} $1/S^0$, i.e. linear spin wave theory (LSWT), is\cite{Anderson52,Kubo52}
\begin{equation}
\varepsilon^\textrm{LSWT}_{\bm k} = \sqrt{4 - \lambda^2 (\cos(k_x) + \cos(k_y))^2}.
\end{equation}
Hence for $\lambda=1$, there are two linearly-dispersing Goldstone modes at the zone center (in sectors $S^z_\textrm{tot} = \pm 1$), consistent with two of three generators of $SU(2)$ being spontaneously broken\cite{Nielsen76}. However, based on general sum rules\cite{Stringari94}, it is known that $|| S^{x,y}_{\bm k} |0\rangle || \sim|\bm k|$ as $\bm k \to \bm 0$, such that the Goldstone modes will have vanishing intensity in the transverse spectral function $\mathcal S^t(\bm k ,\omega)$ at the zone center. Instead, they show up near the ordering wavevector M $= (\pi,\pi)$, since the same sum rules imply an (integrable) divergence $|| S^{x,y}_{\textrm{M}+\bm k} |0\rangle || \sim 1/|\bm k|$ as $\bm k \to \bm 0$.

The first order corrections to the dispersion within spin wave theory are\cite{Oguchi60,Zheng91}
\begin{equation}
\varepsilon^\textrm{LSWT+$1/S$}_{\bm k} = a \; \varepsilon^\textrm{LSWT}_{\bm k} - (1-\lambda^2) \;b \left( \frac{2}{\varepsilon^\textrm{LSWT}_{\bm k}} - \frac{\varepsilon^\textrm{LSWT}_{\bm k}}{2} \right)
\end{equation}
where $a$ and $b$ are ($\lambda$-dependent) constants\footnote{\unexpanded{$a  = 2-\frac{1}{2} \int \frac{\mathrm d \bm q}{4\pi^2} \; \varepsilon_{\bm q}^\textrm{LSWT}$} and \unexpanded{$b = \int \frac{\mathrm d \bm q}{4\pi^2} \; \frac{(\cos(q_x)+\cos(q_y))^2}{\varepsilon_{\bm q}^\textrm{LSWT}}$}, where the integration is over $[-\pi,\pi]\times [-\pi,\pi]$.}. At the isotropic point, this correction is only a momentum-independent rescaling. Higher-order corrections ($1/S^2$ and $1/S^3$) are also known\cite{Hamer92,Zheng93,Syrom10}, which we discuss in section \ref{sec:roton}.

\subsection{Phenomenology of diagonal magnons: a short review}

The purpose of this section is to give a brief (and, unavoidably, partial) overview of some of the salient features which have been the focus of much of the previous work on the excitations of this model.

There is a peculiar property of the LSWT prediction: $\varepsilon_{\bm k}$ is constant along $|k_x| + |k_y| = \pi$. For convenience, we refer to magnons with these momenta as being \emph{diagonal}. This dispersionless feature is a consequence of the more basic fact that at these momenta, the low-order spin wave Hamiltonian vanishes. This also means that the Bogoliubov rotation, which normally mixes the bosons of the two sublattices, is absent there. We thus arrive at the fact that, within LSWT, the diagonal magnons are purely \emph{localized on a single sublattice}.


In fact, this one-dimensional flatness in the spectrum means one has a freedom in choosing a basis of energy eigenstates. By Fourier transforming the momentum eigenstates along one direction, one can thus construct eigenstates which are spatially \emph{localized onto a single diagonal} (of a given sublattice), with alternating signs along this diagonal. In summary, at low order in SWT, diagonal magnons are localized on both a sublattice and a diagonal. (In section~\ref{sec:perturbative}, we show that the same features arise naturally at low order in the Ising expansion.)

Despite being flat in LSWT and LSWT+$1/S$, the diagonal magnons acquire a finite but very small dispersion at higher order\cite{Hamer92,Zheng93,Syrom10,Uhrig13}. Equivalently, this means that magnons can no longer be confined onto a single diagonal. However, it has not yet been investigated whether the aforementioned sublattice-localization persists. We study this both numerically (section~\ref{sec:entanglement}) and perturbatively (section~\ref{subsec:states}).

The SWT predictions at diagonal momenta, $|k_x|+|k_y|=\pi$, do not agree well with other methods or experiments\cite{Ronnow01,Christensen07,DallaPiazza15}---both with respect to single- and multi-magnon features. Examples of previous studies include methods based on quantum Monte Carlo (QMC) combined with analytic continuation\cite{Sandvik01,Shao17}, series expansions in $\lambda$ (up to 14th order)\cite{Singh95,Zheng05}, exact diagonalization  (ED)\cite{Luescher09} and the continuous unitary transform (CUT) \cite{Powalski15,Powalski17}. All these methods predict a more pronounced local minimum of the magnon at $\bm k =\left(\pi,0\right) = X$, referred to as the roton mode, although they do not agree on its exact magnitude or shape (a quantitative discussion is deferred to section~\ref{sec:roton}). More strikingly, they also predict an unusually large weight in the continuum above this local minimum.

The latter phenomenology is also observed in experiment\cite{Christensen07,Headings10,DallaPiazza15}, and exotic scenarios have been given to explain it. For example, it has been argued that near $\bm k \approx X$, the magnon can be seen as two (nearly) deconfined spinons \cite{Ho01,DallaPiazza15,Shao17,Ferrari18}. This interpretation has subsequently been challenged by the CUT method \cite{Powalski15,Powalski17}, which reproduces various salient features based on a picture of strongly-interacting magnons. The intuitive nature of said strong interactions, however, has not yet been clarified. We will argue that an Ising-like domain-wall interaction naturally accounts for it.

\section{Spectral functions} \label{sec:spectral}

In this section we discuss the transverse spectral function $\mathcal S^t(\bm k,\omega)$ as defined in Eq.~\eqref{eq:transverse}. For this, we use the numerical method introduced in Ref.~\onlinecite{Gohlke17}, which we briefly outline here. Firstly, the model~\eqref{eq:model} is put on an infinitely long cylinder whose finite, periodic direction is along a zigzag/staircase path. We define the circumference $L_\textrm{circ}$ in Manhattan distance, i.e. the minimal number of bonds needed to wrap around the cylinder. In this work, $L_\textrm{circ} = 8,10$. The infinite density matrix renormalization group (iDMRG) method \cite{White92,White93,Kjaell13} is used to obtain the ground state\cite{Stoudenmire13}. The dynamical spin-spin correlations $C^{\gamma,\gamma}(\bm r,t)$ can then be calculated by using a matrix-product-operator-based time evolution\cite{Zaletel15}. The spectral function follows directly from Eq.~\eqref{eq:spectral} \cite{Schollwoeck11}.

Let us mention a few technical details before discussing the results. To minimize the effects of Fourier transforming a finite-time window, we use linear prediction\cite{White08} to increase the time window, after which we multiply the data with a Gaussian envelope\footnote{The width of the Gaussian is chosen such that there is only little weight on the data generated by linear prediction.}. This effectively introduces an artificial broadening of the spectral function with full-width-at-half-maximum $2.355 \sigma_\omega$. For a given circumference, we confirm that our results are converged in both bond dimension and inverse time-step by increasing both until the results no longer change. Due to the expensive nature of time-evolving large cylinders, in this work we are limited to bond dimension $\chi \approx 400$ for the largest circumference considered ($L_\textrm{circ}=10$). A typical size that we used for the time-step is $dt = 0.01/J$. The conservation of $S^z_\textrm{tot}$ was implemented explicitly. 

\subsection{Isotropic/Heisenberg model\label{subsec:spectral}}

\begin{figure}
	\includegraphics[trim=0.1cm 0.5cm 0.2cm 0cm]{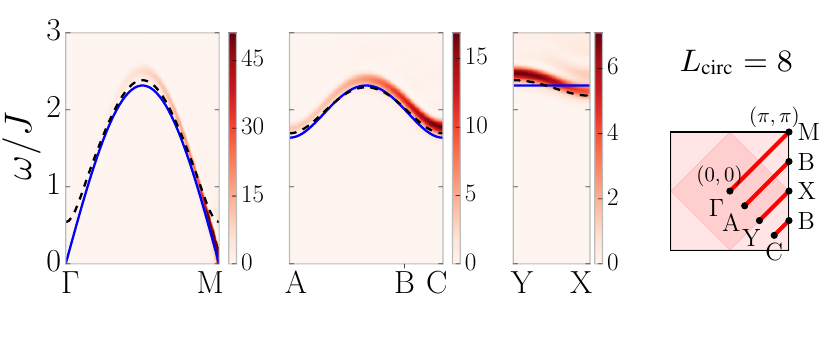}
	\includegraphics[trim=0.1cm 0.5cm 0.2cm 0cm]{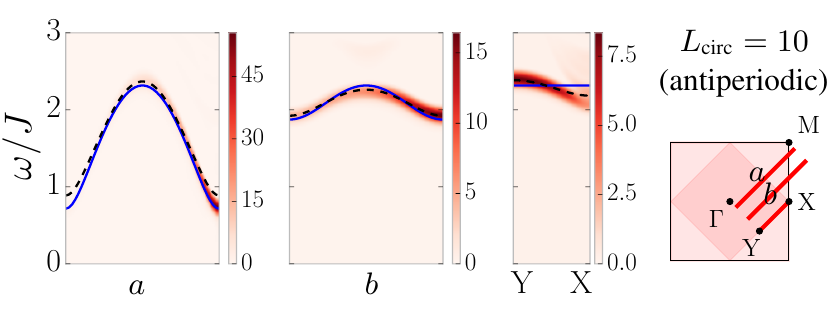}
	\caption{Symmetry-inequivalent momentum cuts of the transverse spectral function for the Heisenberg model ($\lambda=1$). We show results for two infinitely-long cylinders with distinct circumferences $L_\textrm{circ}=8,10$. (The maximum of the color scale is set by the largest value of the spectral function, except along the $\Gamma-\textrm{M}$ line, where we reduced it by $90\%$.) The effective broadening due to the finite time window is $\sigma_\omega \approx 0.055J$ ($L_\textrm{circ} = 8$) and $\sigma_\omega \approx 0.05J$ ($L_\textrm{circ} = 10$). Dashed black line is the series expansion result up to twelfth order evaluated at $\lambda=1$ \cite{Zheng05,Oitmaa18}; solid blue line is LSWT+$1/S$ \cite{Oguchi60,Zheng91}.\label{fig:isotropic}}
\end{figure}

Fig.~\ref{fig:isotropic} shows the transverse spectral function at the isotropic point ($\lambda = 1$). Because of the cylindrical geometry, momentum is discrete along one direction and continuous along the other. This is indicated by the red lines in the Brillouin zone in Fig.~\ref{fig:isotropic}. Since the periodic direction is along a zigzag/staircase path, the momentum cuts are lines of constant $k_x - k_y$. This means we can directly access a line of diagonal magnons (as defined in section \ref{sec:model}), in the figure denoted by the line segment Y--X, where $X = (\pi,0)$ and $Y = \left( \frac{\pi}{2},-\frac{\pi}{2} \right) \cong \left( \frac{\pi}{2},\frac{\pi}{2} \right)$ (by symmetry). In fact, while it is true that $ \left( \frac{\pi}{2},-\frac{\pi}{2} \right)$ and $\left( \frac{\pi}{2},\frac{\pi}{2} \right)$ are symmetry-equivalent in 2D, this is not strictly true on the cylinder geometry. However, such finite-size effects turn out to be small, as discussed in Appendix~\ref{app:analysis}. The same line of diagonal magnons, X--Y, can be accessed for $L_\textrm{circ} = 10$ if there are antiperiodic boundary conditions along the finite direction, shifting the momentum cuts as shown.

We numerically observe the Goldstone modes at the zone center and the ordering wavevector $\textrm{M}=(\pi,\pi)$. Moreover, the intensity vanishes at the zone center, and diverges at M, consistent with the sum rules discussed in section \ref{sec:model}. This agrees with the Goldstone modes predicted by LSWT+$1/S$ (solid blue line), whereas the naive evaluation of the series expansion data (up to $\lambda^{12}$) does not reproduce this\footnote{More sophisticated extrapolation techniques could be and have been used for the low-energy modes near the isotropic point \cite{Singh95,Zheng05}, however in this work we focus on the high-energy modes.} (dashed black line).

On the other hand, along the Y--X line, the series expansion data fares better at reproducing the local minimum at $\textrm{X} = (\pi,0)$. As discussed in section \ref{sec:model}, SWT predicts a flat dispersion along Y-X. Moreover, even in this linear color scale, we can see spectral weight above the single-magnon curve near $\bm k \approx \textrm{X}$. These single- and multi-magnon features at the isotropic point are analyzed in more detail in section \ref{sec:roton}. Here, we limit ourselves to a few general, conceptual remarks.

\begin{figure}
	\includegraphics[trim=0.1cm 0.5cm 0.2cm 0cm]{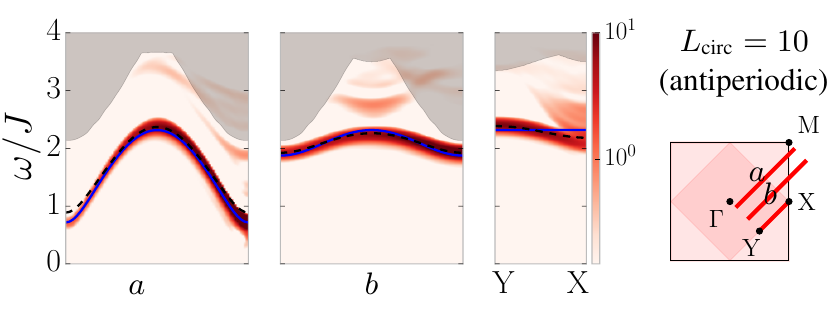}
	\caption{The data of Fig.~\ref{fig:isotropic} (with $L_\textrm{circ}=10$) in log-scale. The shaded region denotes the \emph{kinematic} three-magnon continuum (for this geometry). For $L_\textrm{circ}=10$, this does not start at the one-magnon branch since the momentum cuts do not go through the Goldstone modes. The continuum above the roton mode ($\bm k \approx X$) is outside the kinematic region and it is instead related to (quasi-)bound state physics. \label{fig:isotropic_log}}
\end{figure}

For $L_\textrm{circ} = 8$, the system is gapless at the zone center, hence the multi-magnon continuum starts at the one-magnon branch (up to the miniscule gap introduced by using a finite bond dimension). For the $L_\textrm{circ}=10$ geometry, however, the antiperiodic boundary condition imply that we do not pass through the Goldstone mode, such that the multi-magnon continuum is separated from the one-magnon branch. Nevertheless, these antiperiodic boundary conditions have some useful side-effects. Due to now simulating a gapped system, it is easier to converge the numerics in the bond dimension parameter of the matrix product state describing the ground state. Moreover, it allows the ground state to spontaneously break the symmetry, even at the isotropic point. This is non-trivial given our set-up, since the cylinder is effectively a one-dimensional system (with a large unit cell), such that the Mermin-Wagner-Coleman theorem\cite{Mermin66,Coleman73} should prevent ordering. The catch is that the antiperiodic boundary conditions explicitly break the $SU(2)$ symmetry, although this is not locally noticeable.

This effective gap for $L_\textrm{circ} = 10$ can give us further insight into the physics beyond that of a single magnon. In Fig.~\ref{fig:isotropic_log}, we show the same data in log-scale. We see a continuum right above the single-magnon branch near $\bm k \approx X$. However, this continuum does \emph{not} fall within the frequency region of the kinematic (non-interacting) three-magnon continuum\footnote{Note that due to parity symmetry, the transverse spectral function does not pick up the two-magnon continuum.}. To emphasize this, we have plotted the three-magnon continuum \emph{for this cylinder geometry} in the grey shaded region. Hence, the continuum above the roton mode is instead related to (quasi-)bound states. More precisely, using the insights from the upcoming section~\ref{subsec:evolution}, this continuum is a combination of closely packed three-magnon (quasi-)bound states and a continuum made out of a single magnon and a two-magnon (quasi-)bound state. This is strongly suggestive that the roton mode arises by being repelled from these strongly-interacting states. This agrees with the conclusions of the CUT approach\cite{Powalski15,Powalski17}.

\subsection{Interpolating between the Ising and Heisenberg limits\label{subsec:evolution}}

\begin{figure}
	\includegraphics[scale=1]{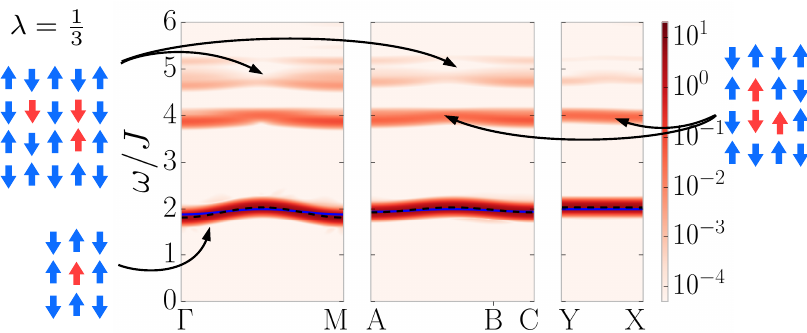}
	\includegraphics[scale=1]{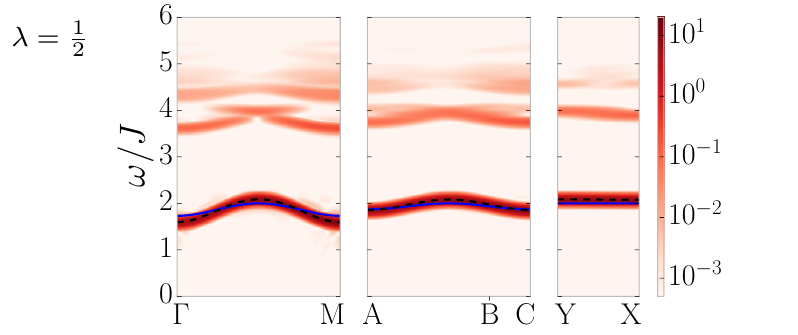}
	\includegraphics[scale=1]{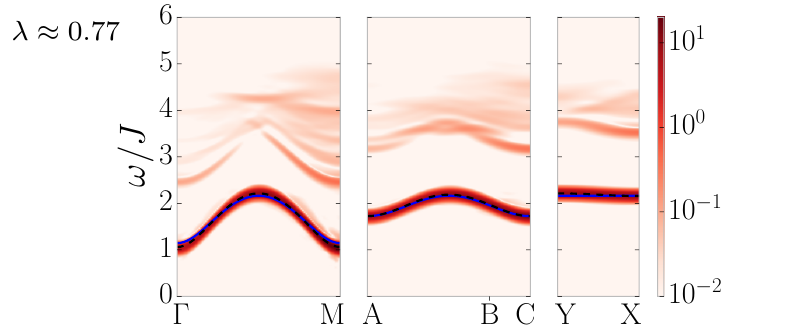}
	\includegraphics[scale=1]{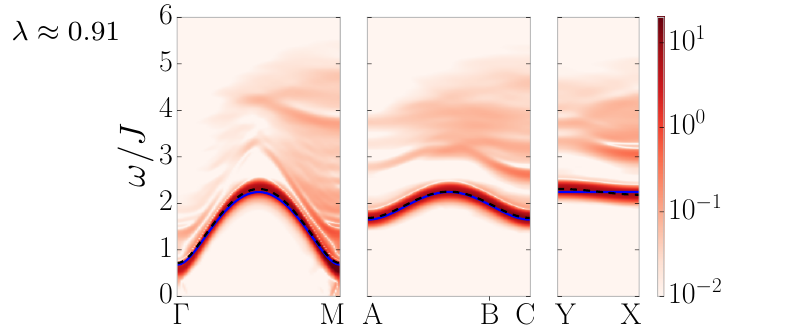}
	\includegraphics[scale=1]{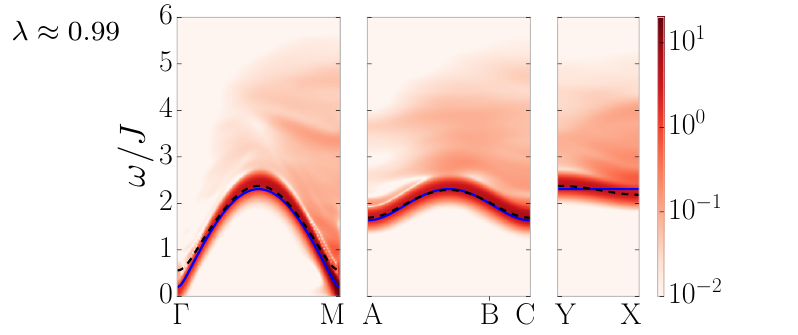}
	\caption{The transverse spectral function as a function of $\lambda$ ($L_\textrm{circ} = 8$, log-scale). Spectral features at the $SU(2)$ point ($\lambda=1$) can be connected back to the Ising limit ($\lambda=0$). Near the Ising limit, we identify the one-magnon branch, three-magnon bound states and three-magnon continuum based on domain-wall counting. Solid blue line is SWT$+1/S$ \cite{Oguchi60,Zheng91}, dashed black line is series expansion\cite{Zheng05,Oitmaa18} up to order $\lambda^{12}$. For the labeling of the momentum directions, see Fig.~\ref{fig:isotropic}. \label{fig:spectral_evolution}}
\end{figure}

Fig.~\ref{fig:isotropic} confirms that the DMRG method can reproduce the roton mode and the strong presence of multi-magnon features near $\bm k \approx X$. However, to gain insight into \emph{how} and \emph{from where} these features appear, it is useful to interpolate from the Ising limit ($\lambda=0$) to the $SU(2)$-symmetric point ($\lambda=1$). This is illustrated in Fig.~\ref{fig:spectral_evolution} for $L_\textrm{circ}=8$. In particular, we demonstrate that the features at the isotropic point can be traced back to those in the Ising limit.

To this end, let us first identify the spectral features close to the Ising limit ($\lambda=\frac{1}{3}$). We can do this by using simple energetic arguments. Note that when $\lambda = 0$, a `magnon' corresponds to a single localized spin flip with energy cost $\frac{J}{2}$ per bond, totaling $2J$. In Fig.~\ref{fig:spectral_evolution}, we see that for for $\lambda = \frac{1}{3}$ the magnon has gained some dispersion, but its energy is still roughly $2J$. Since the transverse spectral function picks up states with $S^z_\textrm{tot} = \pm 1$, the next excitation contains three magnons. Energetically, these magnons prefer to form a bound state whose domain wall counts eight bonds. Indeed, we observe bound states at energy $8 \times \frac{J}{2} = 4J$. There are several such states at this energy due to the internal degree of freedom corresponding to orientation and shape. At even higher energies, there is the kinematic continuum made out of a two-magnon bound state ($6\times\frac{J}{2}$) and a free magnon ($2J$) with total energy around $5J$.

Having identified all spectral features for $\lambda=\frac{1}{3}$, we track their evolution as we tune $\lambda \to 1$ in Fig.~\ref{fig:spectral_evolution}. At $\lambda = \frac{1}{2}$, some of the three-magnon bound states have merged. When $\lambda \approx 0.77$, several of the three-magnon bound states have already been absorbed into the three-magnon continuum. Closer to the isotropic point, $\lambda \approx 0.91$, the three-magnon continuum continues to come down in energy. This trend gradually continues up to $\lambda \approx 0.99$.

We see that the spectral features vary continuously as a function of $\lambda$. In particular, we see that there is no restructuring of the magnon near $\bm k \approx (\pi,0)$ for any $\lambda < 1$. This was a priori not a given. Read backwards, this means that the features near the isotropic point can be continuously traced back to those in the Ising limit. Relatedly, it is worth pointing out that even at the isotropic point, the multi-magnon continuum is not featureless. We discuss this substructure more quantitatively in section~\ref{sec:roton}.

In Fig.~\ref{fig:isotropic_log}, we saw how for $L_\textrm{circ}=10$, there is a continuum above the roton mode which is not made out of kinematic combinations of magnons. In section~\ref{subsec:spectral}, we claimed that it is instead a continuum made up out of (quasi-)bound states. The justification for this claim is that by smoothly decreasing $\lambda$, the observed continuum indeed splits up into three-magnon bound states and a continuum made up out of a magnon and a two-magnon bound state.

\section{Perturbative understanding from the Ising limit} \label{sec:perturbative}

In section~\ref{sec:spectral} we saw that we could connect spectral features of the isotropic model to those near the Ising limit. The purpose of this section is to complement this by gaining insights from low-order perturbation theory in $\lambda$. The point is not to see how well the isotropic point can be described \emph{quantitatively} by a series expansion in $\lambda$ \cite{Singh95,Zheng05}. Instead, we ask what the lowest order processes are that \emph{qualitatively} explain certain features at the isotropic point. Intriguingly, this already naturally leads to a \emph{semi-quantitative} description.

We rewrite Hamiltonian \eqref{eq:model} as $H = H_0 + \lambda V$ with
\begin{equation}
\begin{array}{cl}
H_0 &= J \sum_{\langle\bm n, \bm m \rangle} S^z_{\bm n} S^z_{\bm m}, \\
V &= \frac{J}{2} \sum_{\langle\bm n, \bm m \rangle}  \left( S ^+_{\bm n} S^-_{\bm m} + S^-_{\bm n} S^+_{\bm m} \right).
\end{array}
\end{equation}
The Ising limit $\lambda = 0$ is exactly solvable: the ground state is a product N\'eel state, and the single-magnon excitations consist of localized spin flips. The perturbation $\lambda V$ introduces dynamics to these static excitations. Before going through this in detail, let us give the broad brush strokes to emphasize the simplicity of both the ingredients and results.

\begin{figure}
	\includegraphics[scale=0.45]{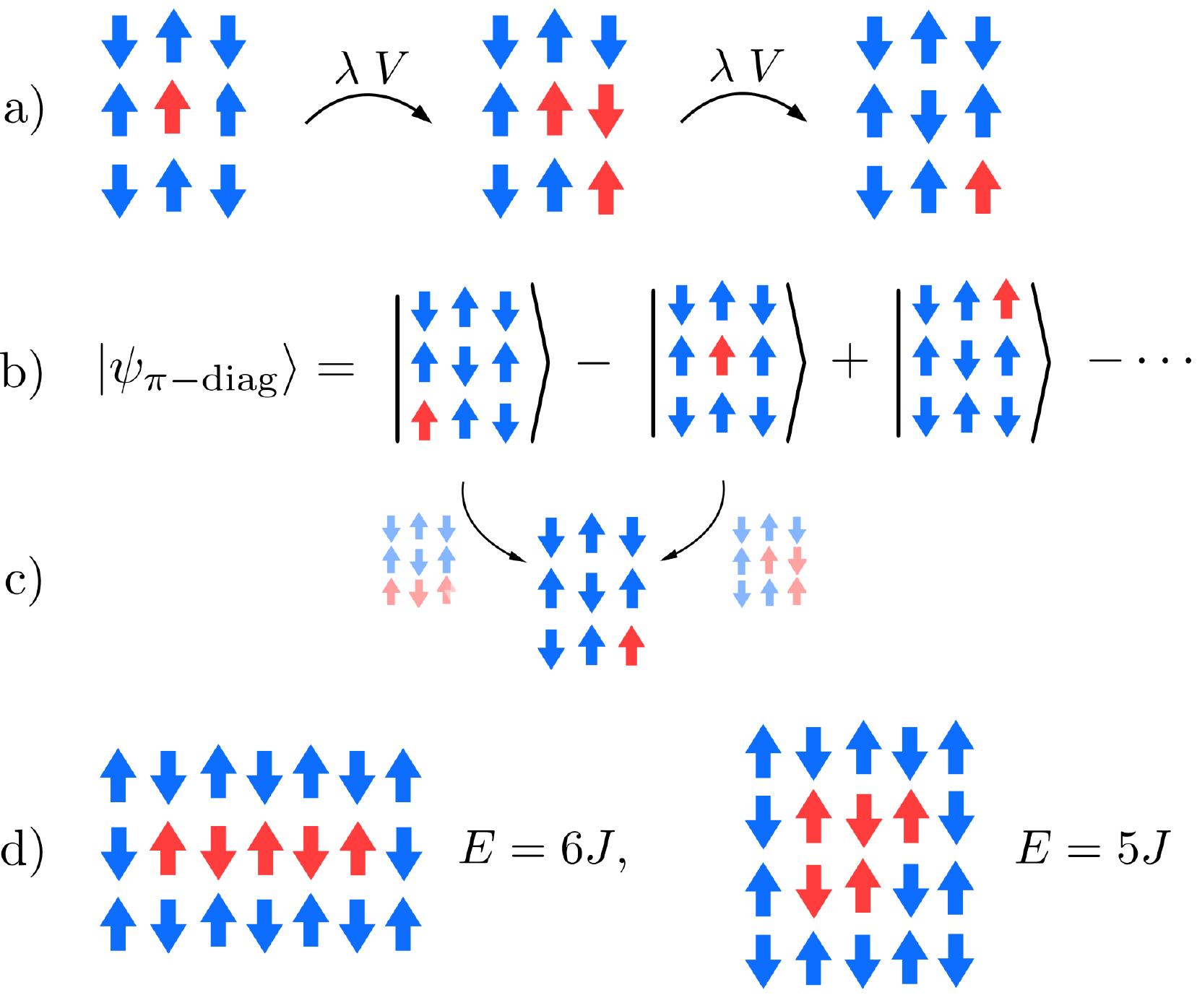}
	\caption{Perturbation theory around the Ising limit. We define the A-(B-)sublattice where spins point down (up) in the ground state. (a) An A-magnon can hop at order $\lambda^2$ through a virtual three-magnon bound state; (b) this one-magnon state which is localized on a single diagonal cannot hop off it at order $\lambda^2$ due to the destructive interference shown in (c); (d) no destructive interference at order $\lambda^4$: different intermediate five-magnon bound states do not have the same energy.\label{fig:pert}}
\end{figure}

\subsection{Overview and summary of the perturbative picture \label{subsec:overview}}

As we will argue, the effective Hamiltonian has contributions at \emph{even} order in $\lambda$ only. The Ising magnons start to hop at order $\lambda^2$ by going through a virtual three-magnon bound state (see Fig.~\ref{fig:pert}(a)). Nevertheless, as we explain in section~\ref{subsec:leading}, magnons with diagonal momentum $|k_x| + |k_y| = \pi$ are still dispersionless at this order. This is equivalent to the statement that if one builds a one-magnon state which is entirely localized on a single diagonal and has momentum $\pi$ along it (see Fig.~\ref{fig:pert}(b)), then it cannot hop off due to destructive interference (see Fig.~\ref{fig:pert}(c)).

The key to the destructive interference traces back to the fact that all three-magnon bound states have the same energy, and hence the virtual paths---half of which come with opposite signs due to the $\pi$-momentum---can cancel exactly. Thus from the viewpoint of the Ising expansion, such destructive interference and the resulting flatness of the diagonal magnons seems accidental. It is hence not surprising that if one goes to next-to-leading order, i.e. $\lambda^4$, the diagonal magnons acquire a dispersion. Indeed, now virtual five-magnon bound states appear, which can have differing energies (see Fig.~\ref{fig:pert}(d)).

Since the emergence of the roton mode is due to the physics of (virtual) bound states, one can indeed say that this phenomenology is due to the attractive interactions between magnons. Since this interaction is so natural in the Ising language, the qualitatively correct physics arises rather easily. Indeed, the resulting dispersion at order $\lambda^4$ does not just correctly reproduce the qualitative features of having a local minimum at $\bm k = (\pi,0)$ and a maximum at $\bm k = \left( \frac{\pi}{2} , \frac{\pi}{2} \right)$, but evaluating it at $\lambda = 1$ even gives a semi-quantitative description for the isotropic model, as we discuss in section~\ref{subsec:nexttoleading}. It is moreover in remarkable proximity to the CUT prediction\cite{Powalski15,Powalski17}, which is a sophisticated framework for strongly-interacting magnons. This success at relatively low order is in contrast with higher-order SWT.

\begin{figure}[h]
	\includegraphics[scale=.34]{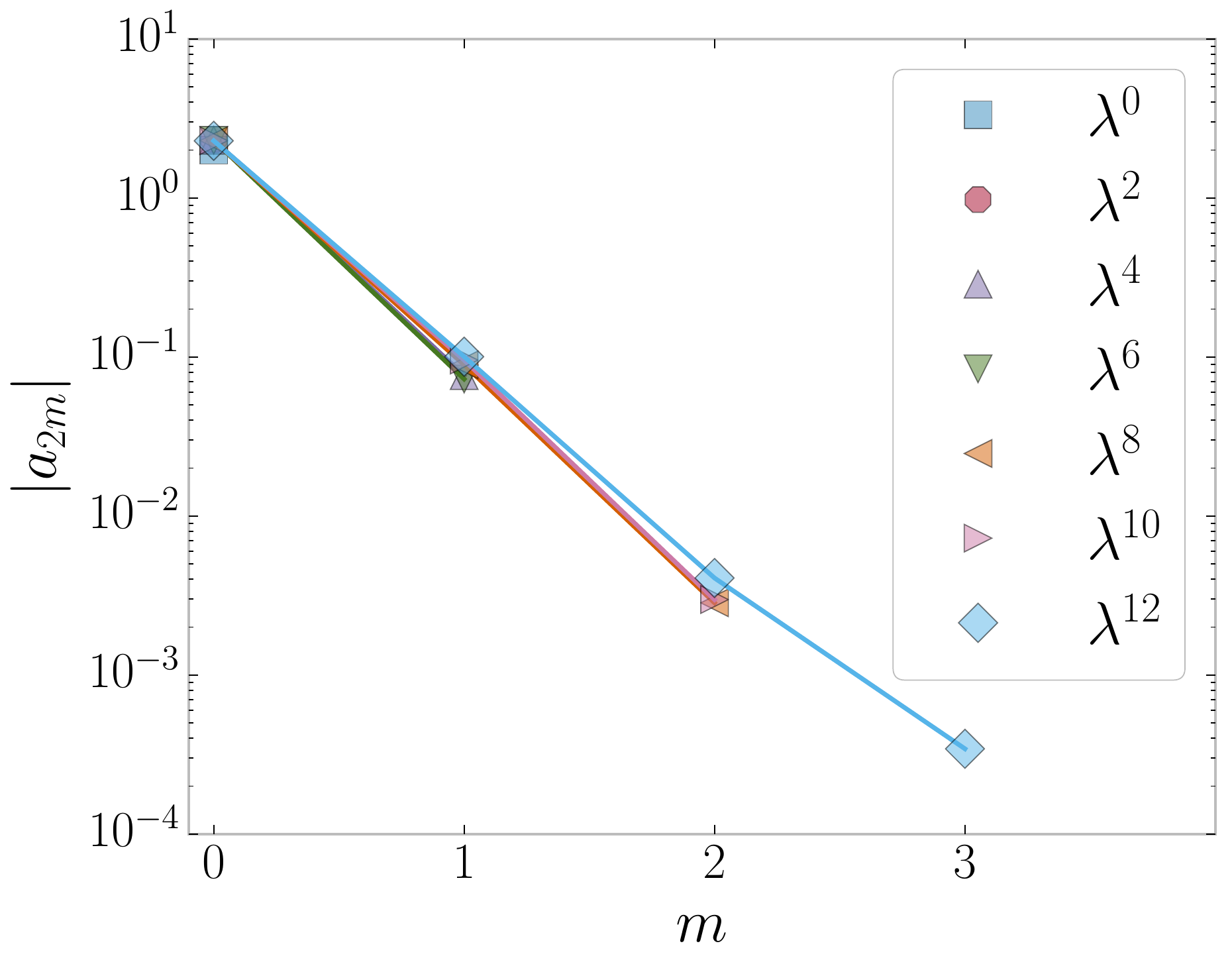}
	\caption{The size of the Fourier coefficients of the dispersion of the diagonal magnons, $\varepsilon_{\pi-k,k} = J \sum_m a_{2m} \cos(2m k)$, obtained from series expansion up to different orders $\lambda^n$, evaluated at $\lambda=1$. This is based on data from Refs.~\onlinecite{Zheng05,Oitmaa18}. We observe that the higher harmonics die off exponentially fast. \label{fig:higherorder}}
\end{figure}

That yet-higher-order corrections don't radically change the physics at hand can be confirmed by repurposing results from previous studies\cite{Singh95,Zheng05}. In Fig.~\ref{fig:higherorder}, we show how the dispersion along the line of diagonal magnons has certain `harmonics' generated at distinct orders in $\lambda^n$ (the first non-trivial harmonic appearing at $\lambda^4$). We see that the higher harmonics die off exponentially fast, justifying a low-order picture. Note that such an exponential decay is a priori not obvious, considering that perturbation theory generically leads to an exponential proliferation of the number of terms.

Lastly, we also consider the perturbed wavefunctions at leading order. In particular, the ground state is dressed by pairs of correlated spin flips, introducing entanglement. This would usually allow one to create magnons associated with one sublattice by acting on the other sublattice. Surprisingly, our perturbative analysis implies that this is \emph{not} possible for diagonal magnons. In other words, they seem to be localized on a given sublattice. This is discussed in section~\ref{subsec:states}. (We also revisit and confirm such sublattice-localization numerically in section~\ref{sec:entanglement}.)

We now provide the details of the story as just described.

\subsection{Ising limit and defining magnons}

In the Ising limit $\lambda = 0$, the ground state is a product N\'eel state. Let us define the A-(B-)sublattice to be where spins point \emph{down} (\emph{up}) in the ground state. For any product state in the spin-$z$ basis, we can count the number of flipped spins on the A-sublattice, relative to the ground state, which we denote by $N_A$ (similarly for $N_B$). Hence, the ground state corresponds to $N_A = 0 = N_B$, whereas the single-magnon states have $N_A = 1$, $N_B = 0$ (called A-magnons) or $N_A = 0$, $N_B = 1$ (B-magnons).

The perturbation $V$ will mix states with different $N_A$ and $N_B$, however $N_A - N_B = S^z_\textrm{tot}$ remains a well-defined quantum number. If we perturb the system such that, for a given momentum, the one-magnon energy scale does not overlap with multi-magnon states, we can \emph{non-perturbatively} label the single-magnon states by $N_A-N_B =1$ or $N_A- N_B = -1$, referred to as A- and B-magnons, respectively. Since $V$ thus cannot connect A-magnons to B-magnons, we can limit our study to A-magnons. More precisely, we are interested in the \emph{effective} Hamiltonian $H_\textrm{eff}$ on $\mathcal H_0$, the Hilbert space of states satisfying $N_A= 1$ and $N_B = 0$.

Note that these Ising magnons are exactly those encountered in Fig.~\ref{fig:spectral_evolution}. As discussed in section~\ref{sec:spectral}, a single magnon has a domain wall crossing four bonds and hence has energy $E_0 = 4\times\frac{J}{2} = 2J$ relative to the ground state.

\subsection{Dispersionless diagonal magnons at leading-order \label{subsec:leading}}

The single-magnon states are completely static and localized in the Ising limit. They moreover stay immobile at first order in $\lambda$. More precisely, denoting the projector onto $\mathcal H_0$ as $P_0$, then at first order we have $P_0 V P_0 = 0$. This is because $V$ creates a pair of A- and B-magnons out of the vacuum: it, e.g., maps $|\uparrow\downarrow\rangle \xrightarrow{V} \frac{J}{2} |\downarrow \uparrow\rangle$, whereas it \emph{annihilates} ferromagnetic bonds. More generally, $V$ flips the \emph{parity} of $N_{A,B}$, hence the conservation of $N_A - N_B$ shows that there are \emph{no} contributions to $H_\textrm{eff}$ at any odd order $\lambda^{2n+1}$.

Thus, by standard perturbation theory, the lowest-order effective Hamiltonian on $\mathcal H_0$ is
\begin{equation}
H_\textrm{eff} = E_0 P_0 + \lambda^2 P_0 V G_0 V P_0 + \mathcal O(\lambda^4),
\end{equation}
where $G_0 = (E-H_0)^{-1}$. This indeed introduces hopping, as shown in Fig.~\ref{fig:pert}(a): the A-magnon can hop to any of the eight nearest sites (on the same sublattice) by going through a virtual three-magnon bound state. As discussed in section~\ref{sec:spectral}, such a bound state involves eight \emph{ferromagnetic} bonds, with total energy cost $8 \times \frac{J}{2} = 4J$---whereas the cost of a single magnon is $E_0 = 2J$. Thus, the path shown in Fig.~\ref{fig:pert}(a) carries a weight $\frac{\lambda J}{2} \times \frac{1}{2J-4J} \times \frac{\lambda J}{2} = -\lambda^2 \frac{J}{8} $.

Magnons can thus hop at order $\lambda^2$. However, certain superpositions are immobile due to destructive interference. Consider, for example, the state shown in Fig.~\ref{fig:pert}(b): it is localized on a single A-diagonal, with an alternating sign structure. (We can say its momentum along the diagonal is $\pi$.) The magnon is unable to hop off the diagonal at order $\lambda^2$. This is illustrated in Fig.~\ref{fig:pert}(c), showing two destructively interfering paths. It is important that both paths go through a virtual three-magnon bound state (both with energy $4J$), such that the two weights cancel exactly. Equivalently, all A-magnon momentum eigenstates with $|k_x| + |k_y| = \pi$---which we referred to as diagonal magnons in the previous sections---have constant energy. In summary, the diagonal magnons are dispersionless in the Ising expansion up to order $\mathcal O(\lambda^4)$. It is interesting to note that this coincides with the LSWT$(+1/S)$ predictions in section~\ref{sec:model}.

\subsection{Roton mode at next-to-leading order \label{subsec:nexttoleading}}

It is hence important to see what happens at sub-leading order in $\lambda$. Here we follow the perturbative scheme by Kato\cite{Kato49} and Takahashi\cite{Takahashi77}. In terms of $H_0$ and $V$, they constructed a general-purpose unitary mapping $\Gamma_\lambda: \mathcal H_0 \to \mathcal H$ which embeds the unperturbed states into the space spanned by the true eigenstates (the latter being $\lambda$-dependent). This object gives us access to the effective Hamiltonian
\begin{equation}
H_\textrm{eff} := \Gamma_\lambda^\dagger H \Gamma_\lambda =: \sum_{n=0}^\infty \lambda^n H_\textrm{eff}^{(n)}.
\end{equation}
As argued before, $H_\textrm{eff}^{(2n+1)} = 0$. Moreover, $H_\textrm{eff}^{(0)} = E_0 P_0$ and $H_\textrm{eff}^{(2)} = P_0 V G_0 V P_0 $. From knowing the aforementioned object $\Gamma_\lambda$ (which, for completeness, we reproduce as a function of $H_0$ and $V$ in Appendix~\ref{app:pert}), one can derive that
\begin{equation}
H_\textrm{eff}^{(4)} = P_0 V \tilde G_0 V \tilde G_0 V \tilde G_0 V P_0 - \frac{1}{2} \left\{ H^{(2)}_\textrm{eff}, P_0 V \tilde G_0^2 V P_0 \right\}
\end{equation}
where $\tilde G_0 = Q_0 G_0 Q_0 = Q_0 (E-H_0)^{-1} Q_0$ and $Q_0 = 1-P_0$.

From this, one can calculate that the diagonal magnons acquire a dispersion at this order. The reason for this is in fact simple, as hinted at in section~\ref{subsec:overview}. Fig.~\ref{fig:pert}(d) shows two possible virtual five-magnon bound states that can appear as intermediate states. These two states have domain walls extending over, respectively, twelve and ten bonds. Their energy is thus different, and one should not expect perfect destructive interference.

More precisely, the resulting dispersion at order $\lambda^4$ is described by a simple cosine-like dispersion for the diagonal magnons: $\varepsilon_{\pi-k,k} =a + b\lambda^2 - c\lambda^4 (d+ \cos(2k)) + \mathcal O(\lambda^6)$ with $a,b,c,d>0$. This has a local (roton) minimum at $\bm k = (\pi,0)$ and a maximum at $\bm k = \left( \frac{\pi}{2},\frac{\pi}{2} \right)$. Moreover, evaluating this at $\lambda = 1$ already gives a semi-quantitative description of the isotropic model. We refer the reader interested in quantitative details to section~\ref{sec:roton}, which is devoted to an in-depth comparison between various different methods.

In summary, the roton mode naturally appears in the Ising expansion. Through the property of all three-magnon bound states having the same energy, the local minimum at $\bm k = (\pi,0)$ is absent at leading order, already indicating that it is less pronounced at the isotropic point. At the same time, its salient features readily appear at next-to-leading order, leading to a semi-quantitatively correct description. From this point of view, it is indeed an interacting-magnon effect, where the interaction is based on a simple domain-wall counting in the Ising limit.

\onecolumngrid

\begin{figure}[h]
	\includegraphics[scale=1]{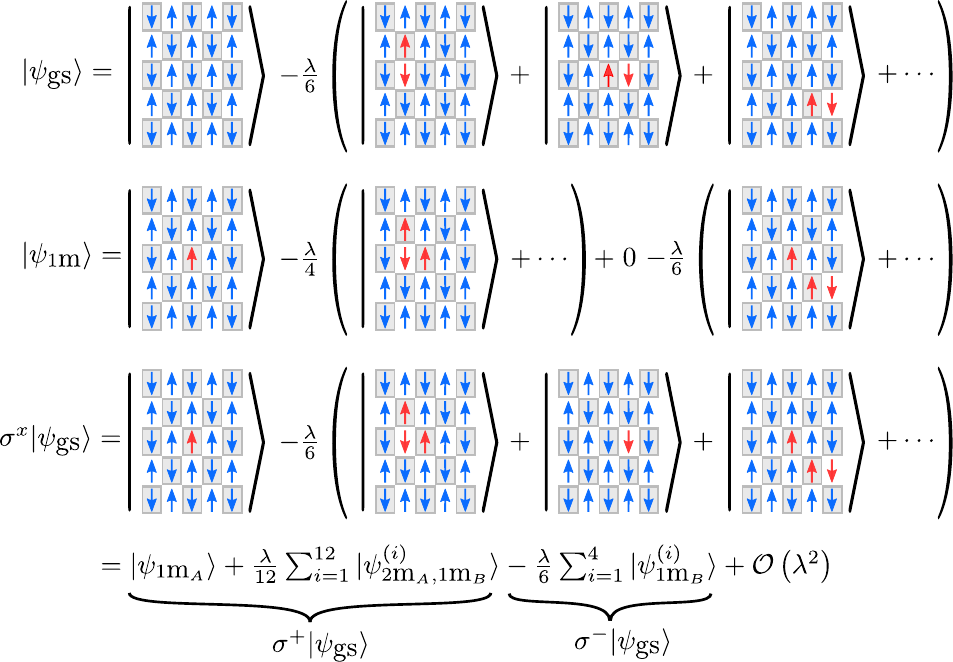}
	\caption{Dressing of the Ising limit ground state and a single A-magnon at leading order in perturbation theory. The gray boxes denote the A-sublattice. The entanglement structure of the above states can be used to understand why the diagonal magnons seem to be localized on a given sublattice, as discussed and observed numerically in section~\ref{sec:entanglement}. \label{fig:perturbation_states}}
\end{figure}
\twocolumngrid

\subsection{Sublattice-localization of diagonal magnons \label{subsec:states}}

Aside from looking at the effective Hamiltonian, it can be instructive to consider how the eigenstates evolve with $\lambda$. This is exactly the information encoded in $\Gamma_\lambda$. In Fig.~\ref{fig:perturbation_states} we show the leading-order results, both for the ground state as well as a localized A-magnon. (One would have to Fourier transform the latter to obtain an energy eigenstate.) We see that these states are dressed with `pair fluctuations' whilst staying within a well-defined $N_A - N_B =\sum S^z_\textrm{tot} = 0,1$ sector.

Having access to the perturbed states, we can ask what excitations are created upon acting with a local operator on the ground state. Fig.~\ref{fig:perturbation_states} shows that at this order, a $\sigma^x$ operator does not just create a single magnon, but also three-magnon bound states. This is to be expected and is directly in line with the spectral weight observed in Fig.~\ref{fig:spectral_evolution}.

It is more interesting to consider what happens when applying $\sigma^-$ on the A-sublattice. This brings us into the sector $N_A - N_B = -1$. In other words, by acting with this operator on the A-sublattice, we create a B-magnon. This is not possible in the product state Ising limit $\lambda \to 0$, where acting with $\sigma^-$ on the A-sublattice annihilates the ground state. But as shown in Fig.~\ref{fig:perturbation_states}, the perturbation $V$ introduces entanglement, such that $\sigma^-_{\bm n} |\psi_\textrm{gs}\rangle$ is nonzero and has a B-magnon on the four B-sites adjacent to the original site $\bm n \in A$.

However, something unusual happens for diagonal momenta. Note that for any given B-site, there are four adjacent A-sites. If the operator we acted with on the A-sublattice has momentum $|k_x| + |k_y| =\pi$, then half of these four sites carry a positive sign, and half a negative sign, so that there would be perfect cancellation. In other words: we are not able to create a B-magnon with $|k_x| + |k_y| = \pi$ by acting on the A-sublattice.

The above used an explicit calculation, but the essential mechanism at play should hold at all orders. If we act on a given site of the A-sublattice, then by the $90^\circ$ symmetry of the model, the signs and weights will be the same in all four directions. However, this is incompatible with the alternating sign structure of diagonal momenta. This argument suggests that as long as we act on a single site of the A-sublattice, we \emph{cannot} create a B-magnon with a diagonal momentum. We investigate this claim of \emph{sublattice-localization} non-perturbatively in the following section, detailing its relationship to entanglement.



\section{Entanglement and sublattice-localization of diagonal magnons} \label{sec:entanglement}

\begin{figure}
	\includegraphics[scale=1]{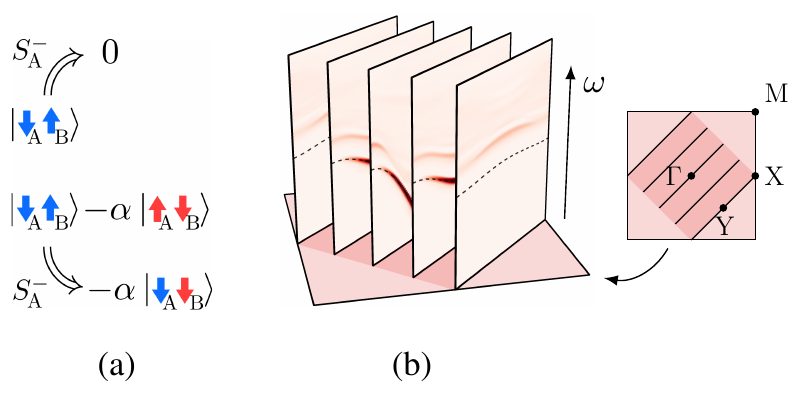}
	\caption{Sublattice-localization of diagonal magnons. (a) If there is entanglement in the ground state (due to pair fluctuations), then it is generically possible to create a B-magnon by acting on the A-sublattice. (b) The \emph{sublattice spectral function} $\mathcal S_{A\to B}(\bm k, \omega)$ obtained with DMRG is shown: this measures whether acting on the A-sublattice can create B-magnons. As defined in Eq.~\eqref{eq:S_ent}, we act with $S^-$ exclusively on the A-sublattice (where the expectation value already points down). The ground state entanglement is responsible for the non-zero spectral weight on the single-magnon branch (dashed line). Surprisingly, there is \emph{no} weight on the diagonal magnons (edge of shaded square). The plot is for $\lambda = 0.95$ and $L_\textrm{circ}=8$. The same conclusion seems to hold when acting with multi-site operators. \label{fig:no_entanglement}}
\end{figure}

In this section we discuss a peculiar property of the entanglement in this model.
As before, we define the A-sublattice where the spins point down in the ground state (opposite for the B-sublattice). The ground state is in the sector $S^z_\textrm{tot} = 0$, and a magnon associated with the B-sublattice, a \emph{B-magnon}, is in the sector $S^z_\textrm{tot} = -1$.

If the ground state had \emph{no} entanglement between the two sublattices, then by acting on the A-sublattice, one could \emph{not} create a B-magnon. The intuitive idea is sketched in Fig.~\ref{fig:no_entanglement}(a), but the more precise wording is as follows: if $|\Psi_\textrm{gs}\rangle = |\psi_A\rangle \otimes |\psi_B\rangle$ (where $|\psi_{\alpha = A,B}\rangle$ lives on the $\alpha$-sublattice), then, e.g., $S^-_{\bm n}$---which puts us in the $S^z_\textrm{tot}$ sector of a B-magnon---\emph{annihilates} the ground state if $\bm n \in A$. To argue this, note that since the ground state is an eigenstate of $S^z_\textrm{tot}$, then the factorization implies that $|\psi_{\gamma = A,B}\rangle$ must be an eigenstate of $S^z_{\gamma =A,B} := \sum_{\bm n \in \gamma} S^z$. Moreover, since the product N\'eel state has a finite\footnote{For overlaps to be well-defined, one can apply the argument to finite systems.} overlap with $|\Psi_\textrm{gs}\rangle$, this fixes the eigenvalue of $S^z_A$ to be as (algebraically) small as possible. Hence, acting with the lowering operator on A must annihilate the state.

The actual ground state will of course have entanglement; for example, at leading order in $\lambda$, there are two-site spin-flips (`pair fluctuations') which entangle the two sublattices, see e.g. Fig.~\ref{fig:no_entanglement}(a) or Fig.~\ref{fig:perturbation_states}. Thus, it is indeed generically possible to create a B-magnon by acting on the A-sublattice. However, this does not seem to be true for diagonal magnons, i.e. when $|k_x| + |k_y| = \pi$. To make this precise, we introduce what we call the \emph{sublattice spectral function},
\begin{equation}
\mathcal S_{A \to B}(\bm k, \omega) =  \sum_\alpha \delta(\omega - (\omega_\alpha-\omega_0)) \; |\langle \alpha | \tilde S^-_{A,\bm k} |0\rangle|^2 \label{eq:S_ent}
\end{equation}
where $\tilde S^-_{A,\bm k} = \sum \limits_{\bm r \in A} e^{i \bm{k\cdot r}} S^-_{\bm r}$. There are two crucial differences that distinguish it from the usual transverse spectral function as in Eq.~\eqref{eq:spectral}. Firstly, we only act on the A-sublattice, where spins point down in the symmetry-broken ground state. Secondly, we act with the lowering operator $S^-$, putting us in the $S^z_\textrm{tot}$ sector of B-magnons.

In Fig.~\ref{fig:no_entanglement}, we show this sublattice spectral function $\mathcal S_{A \to B}(\bm k,\omega)$. For convenience, we consider $\lambda = 0.95$ instead of the isotropic point, as this allows us to tell the one-magnon branch straightforwardly apart from the multi-magnon sector. We see the response is non-zero on almost the whole single-magnon branch (dashed lines). However, the spectral weight is exactly zero for any of the diagonal magnons (i.e. along the border of the shaded region in the Brillouin zone). We conclude that the diagonal magnons appear to be localized on their respective sublattices.

In section~\ref{subsec:states} we gave a symmetry-based argument for the sublattice-localization within the perturbative framework, using the fact that we act with a single-site operator. However, we have also numerically confirmed that the same absence of spectral weight occurs even if we act with \emph{multi-site} operators localized on the A-sublattice (not shown). We have not found an explanation for this and it would be interesting to study this in more detail. It is an open question whether there is a probe that could directly access $\mathcal S_{A \to B} (\bm k, \omega)$ in an experimental set-up.

\section{Quantitative analysis at diagonal momenta} \label{sec:roton}

In this section, we analyze the roton mode at $\bm k = (\pi,0) \cong X$ and its associated multi-magnon features in more quantitative detail, including a comparison to previous work.

\subsection{Depth of the anomalous mode at \texorpdfstring{$\bm{k = (\pi,0)}$}{$k = (\pi,0)$} \label{subsec:amplitude}}

\begin{figure}[h]
	\includegraphics[width=0.85\linewidth]{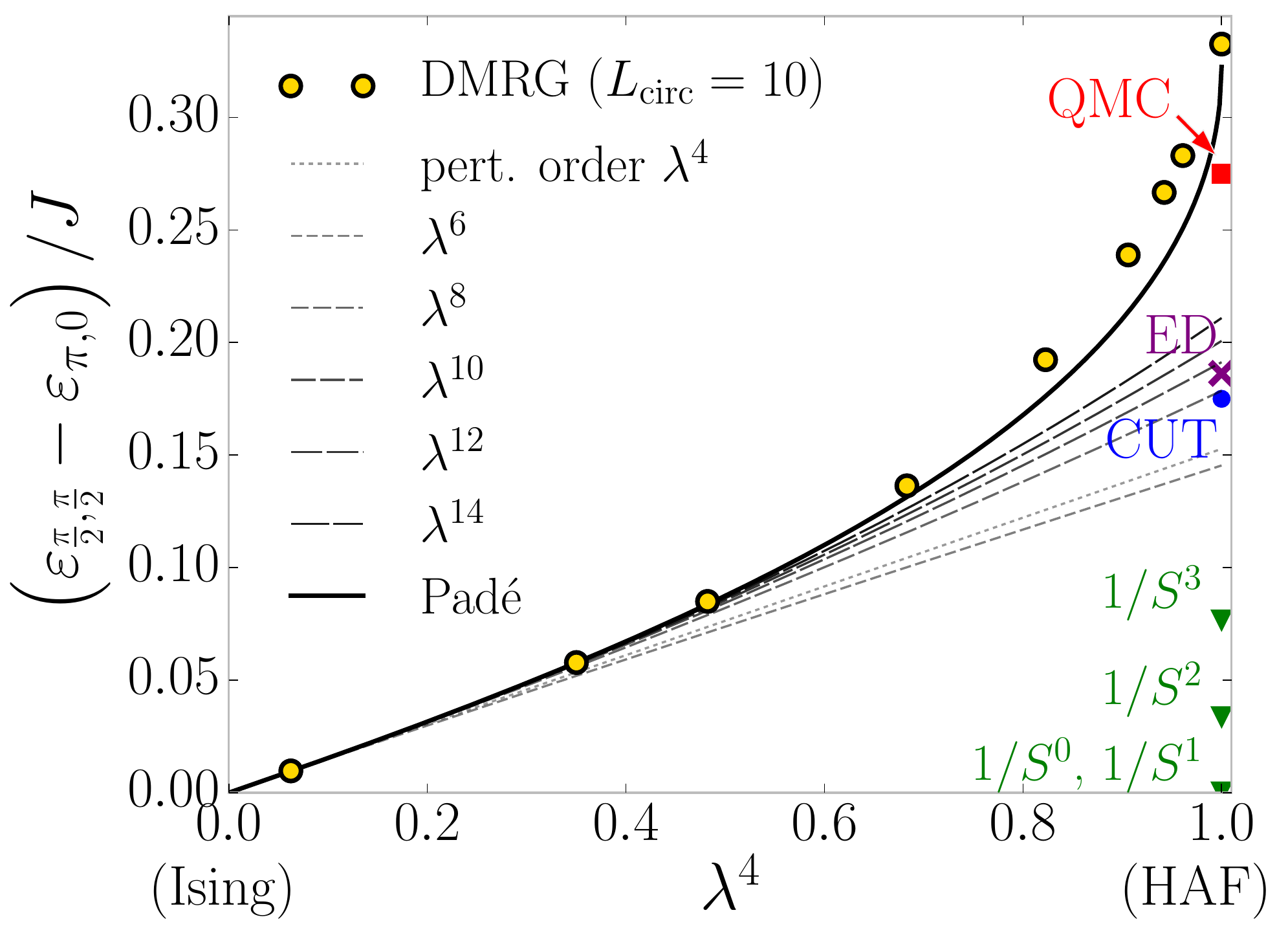}
	\caption{Depth of the roton mode as a function of $\lambda$. Yellow dots are the DMRG results ($L_\textrm{circ}=10$). Dashed lines are series expansion results to various orders\cite{Singh95,Zheng05}. We obtain the solid line by rewriting series expansions in terms of $\delta = 1-\sqrt{1-\lambda^2}$ and subsequently making a Pad\'e approximant (details in main text). At the isotropic point ($\lambda=1$), we compare to QMC\cite{Sandvik01,Shao17}, CUT\cite{Powalski15,Powalski17}, ED\cite{Luescher09} and SWT\cite{Anderson52,Kubo52,Oguchi60,Zheng91,Hamer92,Zheng93,Syrom10}. \label{fig:roton_evolution}}
\end{figure}

In Fig.~\ref{fig:roton_evolution}, we consider the depth of the roton mode, i.e. the maximum of the dispersion at $\bm k = \left( \frac{\pi}{2}, \frac{\pi}{2} \right) \cong Y$ relative to the local minimum at $\bm k = (\pi,0) \cong X$. This is shown as a function of the parameter $\lambda$. As derived in section~\ref{sec:perturbative}, we expect the dispersion to scale as $\sim \lambda^4$ for small $\lambda$. For that reason, we scale our axis accordingly. The numerical results obtained with our DMRG-based method for $L_\textrm{circ}= 10$ (up to $\chi=400$) are shown as yellow dots. At the isotropic point ($\lambda=1$), we plot the predictions of QMC\cite{Sandvik01,Shao17} (extrapolated from up to $N =48 \times 48$), CUT\cite{Powalski15,Powalski17}, ED\cite{Luescher09} (extrapolated from up to $N = 36$) and SWT\cite{Anderson52,Kubo52,Oguchi60,Zheng91,Hamer92,Zheng93,Syrom10}.

This quantity is extracted from the spectral function by fitting the single-magnon response with a Gaussian\footnote{Note that the width of the Gaussian is known; see section~\ref{sec:spectral}.}. Near the isotropic point, this fitting is somewhat subtle, as the one-magnon response is not easily separated from the multi-magnon weight. Fitting the (quasi-)bound states just above the single-magnon branch---which are highly relevant near $\bm k \approx (\pi,0)$---as well, we obtain results which are stable with respect to the numerical parameters.

At the isotropic point, the method that DMRG is closest to is QMC. For $L_\textrm{circ} = 10$, we obtain $\varepsilon_{\frac{\pi}{2},\frac{\pi}{2}} \approx 2.40J$ and $\varepsilon_{\pi,0} \approx 2.06J$-$2.07J$. This can be compared to the QMC\cite{Shao17} extrapolations $\varepsilon_{\frac{\pi}{2},\frac{\pi}{2}} \approx 2.40J$ and $\varepsilon_{\pi,0} \approx 2.13J$. We are unable to perform a finite-circumference analysis, since for $L_\textrm{circ} = 6$ there are domain-wall excitations (wrapping around the circumference), while for $L_\textrm{circ} = 8$ the system is gapless such that we expect stronger finite-circumference effects. However, one can compare our results to the finite-size QMC data with linear dimension $10$, corresponding to our largest cylinder circumference. In that case, QMC obtains\cite{Shao17} $\varepsilon_{\pi,0} \approx 2.2 J$. Taking this to give a rough finite-size estimate, we note that we are within the same distance to the extrapolated QMC data (although at the opposite extreme).

It is illuminating to not just focus on the isotropic case, but to track the evolution as a function of $\lambda$ as shown in Fig.~\ref{fig:roton_evolution}. The dashed lines are from series expansions\cite{Singh95,Zheng05} to different orders in $\lambda$. If we track the lowest-order result $\sim \lambda^4$ toward the isotropic point $\lambda =1$, we already obtain the correct order of magnitude. This is moreover in striking proximity to the CUT prediction. As we include higher order terms, we see that the dispersion gradually creeps up, showing no real sign of convergence. However, since SWT results are analytic in the modified parameter $\delta = 1-\sqrt{1-\lambda^2}$, it is suggestive to rewrite the series expansion in terms of $\delta$. Doing so, and building a Pad\'e approximant out of it, we obtain the solid line in Fig.~\ref{fig:roton_evolution}.

We find that the Pad\'e approximant is remarkably robust: the approximants $[3,3]$, $[4,2]$, $[5,2]$, $[4,3]$, $[3,4]$ all give virtually indistinguishable results! This stability suggests that the solid line could be a reasonable prediction for the true evolution of the dispersion as a function of $\lambda$. Exactly \emph{at} the isotropic point, there is a small caveat: any power series in $\lambda$, when rewritten in terms of $\delta$, generically predicts a diverging slope at $\delta = 1$. Hence, also in this case, we find that the dashed curve is finite at $\delta = 1$ but its slope is vertical. It is not clear whether this particular feature is physical or not. Other than that, we expect that the Pad\'e approximant should be reliable and we are encouraged by the fact that our numerical results agree so well with the Pad\'e curve, indicating that finite-size corrections for $L_\textrm{circ} = 10$ are already rather small.

It would be interesting to investigate to what extent the Pad\'e approximant gives the correct prediction. In particular, it might be worthwhile to test and compare the other methods\footnote{We note that applying the CUT method developed in Refs.~\onlinecite{Powalski15,Powalski17} to $\lambda < 1$ would be distinct from the CUT method that was applied to the XXZ model in Ref.~\onlinecite{Dusuel10}: in the latter case, the CUT method was perturbative in the parameter $\lambda$, hence being an alternative way of calculating  the series expansion coefficients.} at $0 < \lambda <1$. 

\subsection{Dispersion relation}

\begin{figure}
	\includegraphics[scale=.35]{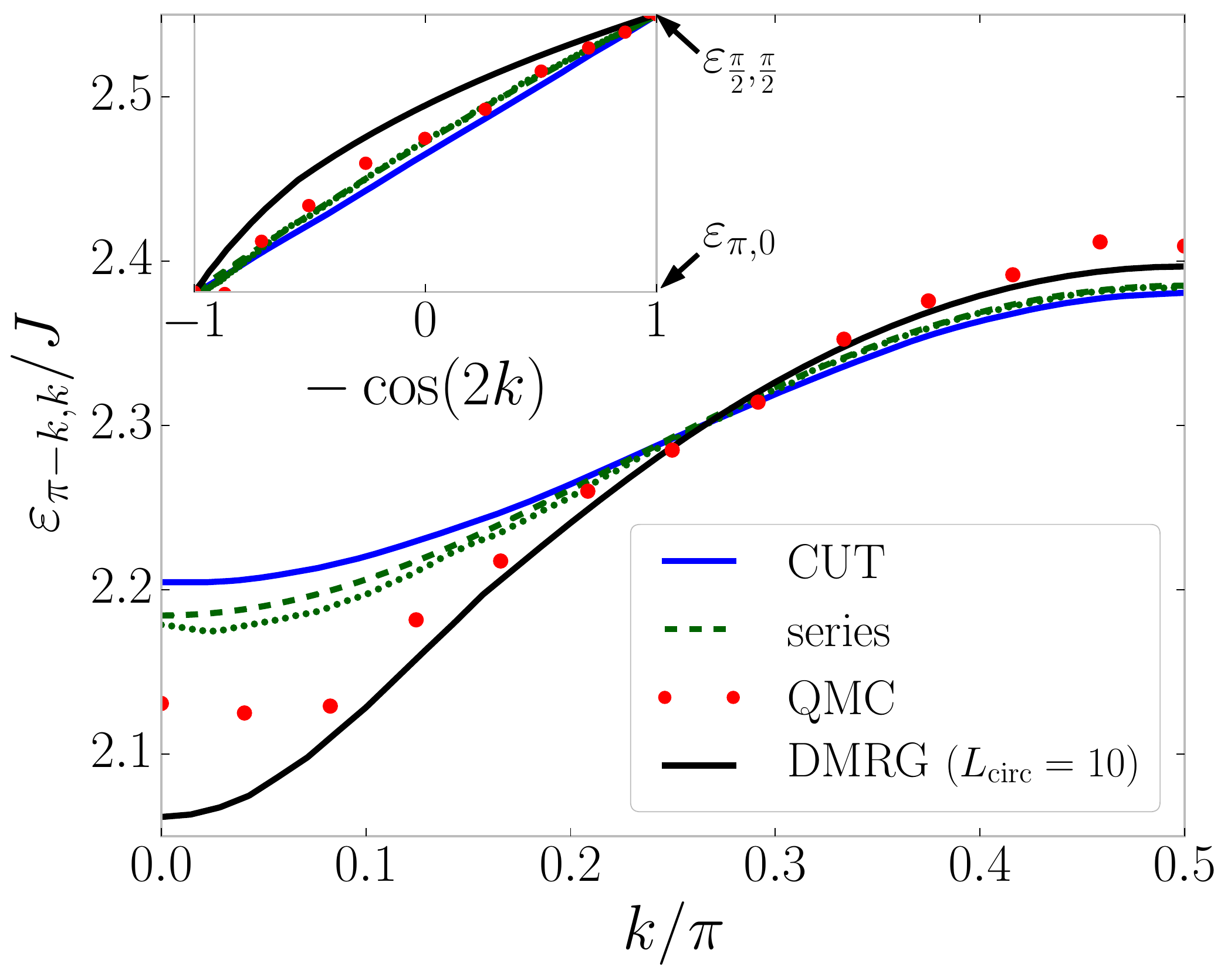}
	\caption{Dispersion of the diagonal magnons. The solid black line is the DMRG result ($L_\textrm{circ}=10$). The green lines are the series expansion results: the dashed line\cite{Zheng05,Oitmaa18} is the result to order $\lambda^{12}$ evaluated at $\lambda=1$, whereas the dotted green line is extracted from the plot in Ref.~\onlinecite{Zheng05}. We also compare to QMC\cite{Sandvik01,Shao17} and CUT\cite{Powalski15,Powalski17}. Inset: rescaled dispersions to compare functional forms. \label{fig:roton_comparison}}
\end{figure}

Aside from studying the depth of the roton mode, we also consider its shape. Our numerical result is shown in Fig.~\ref{fig:roton_comparison} (solid black line). We compare it to the functional forms obtained by CUT, QMC and series expansion. As discussed in section~\ref{sec:perturbative}, the lowest-order non-trivial prediction from the Ising expansion is a simple cosine-like dispersion. This is not significantly altered at higher orders in $\lambda$ (or at the very least, only very slowly so), as was shown in Fig.~\ref{fig:higherorder}. The purpose of the inset is to show a comparison with this simple cosine.

Remarkably, the CUT result is perfectly fit by a single harmonic. In conjunction with section~\ref{subsec:amplitude}, we thus conclude that the fourth order extrapolation from the Ising expansion seems to be in striking proximity to the CUT prediction of the roton dispersion. Our result, on the other hand, while being dominated by the same cosine, also contains the higher harmonics (which are qualitatively generated in the Ising expansion). We point out that the QMC curve has more structure near $\bm k \approx X$, with a possible small subsidiary maximum at the X point itself; it would be interesting to investigate its origin.

\subsection{Multi-magnon features}

\begin{figure}
	\includegraphics[scale=0.3]{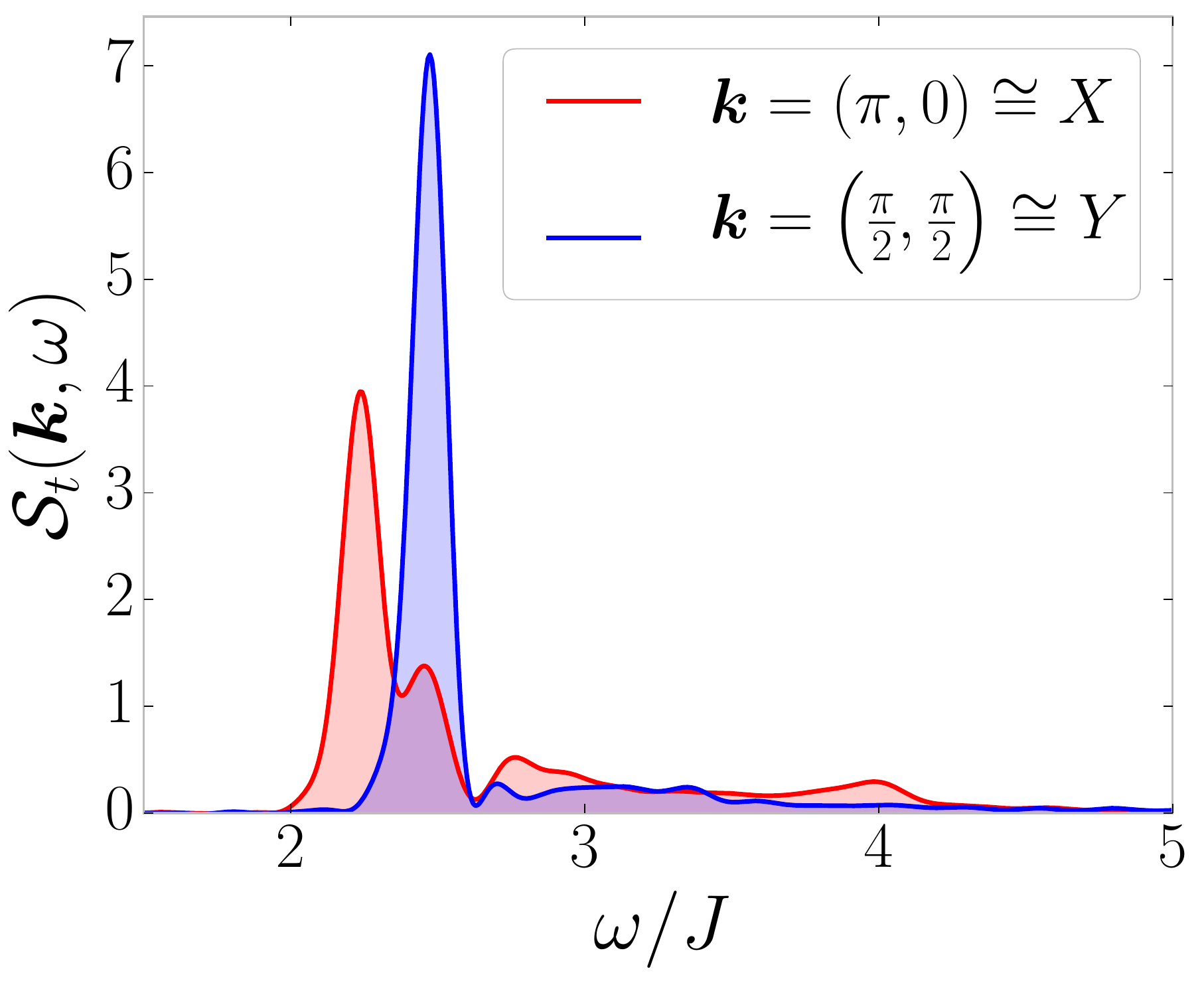}
	\caption{Transverse spectral function $\mathcal S^t(\bm k, \omega)$ for the Heisenberg model ($\lambda = 1$) with $L_\textrm{circ} = 8$ and broadening $\sigma_\omega \approx 0.055J$. The second peak for $\bm k = (\pi,0) \cong X$ is the three-magnon (quasi-)bound state.\label{fig:spectralcut}}
\end{figure}

Lastly, in Fig.~\ref{fig:spectralcut} we show a more detailed slice of the transverse spectral function $\mathcal S^t(\bm k,\omega)$ first shown in in Fig.~\ref{fig:isotropic}. At two values of the momentum, $\bm k = (\pi,0) \cong X$ and $\bm k = \left( \frac{\pi}{2} , \frac{\pi}{2} \right) \cong Y$, we show the spectral weight as a function of $\omega$. This numerical data is for the smaller circumference $L_\textrm{circ} = 8$, since as discussed in section~\ref{sec:spectral}, in that case the system is gapless. Being gapless, one expects there to be more significant finite-size effects on a \emph{quantitative} level, but the \emph{qualitative} shape should look more like the 2D limit than the $L_\textrm{circ}=10$ data would.

For either momentum, we clearly see the single-magnon peak (broadened due to our finite-time window, as explained in section~\ref{sec:model}) and a broad three-magnon continuum. Moreover, for $\bm k = (\pi,0) \cong X$, we recognize a second, smaller peak. This is a three-magnon (quasi-)bound state.

We would like to comment on the following two features of Fig.~\ref{fig:spectralcut}. Firstly, there is considerable weight in the multi-magnon sector at $\bm k = (\pi,0)$, and not at $\bm k = \left( \frac{\pi}{2},\frac{\pi}{2} \right)$. Due to not having a tight grasp on finite-size effects for $L_\textrm{circ}=8$, we do not believe there is great value in quoting precise numbers, but at $\bm k = (\pi,0)$, only roughly half the weight is on the single magnon. Secondly, there is certainly a substructure to the multi-magnon weight. This seems to be in contrast to the featureless spectral function observed in a recent Monte Carlo study\cite{Shao17}, which however considers the sum of the transverse and longitudinal spectral function. We do not expect the longitudinal contribution to completely smear out the substructure; in fact, the CUT analysis indicates the presence of strong resonances in the latter\cite{Powalski15,Powalski17}. Due to the absence of a finite-size analysis, the substructure we observe is not conclusive and it would be interesting to investigate this further.

\section{Conclusion} \label{sec:conclusion}

We have studied the spectral properties of the Heisenberg model, allowing for Ising anisotropy, using two complementary methods: a DMRG-based approach to obtain the unbiased dynamical structure factor (for certain circumferences), and a low-order perturbative expansion around the Ising limit to give insights into the physical mechanisms at play.

One of our key messages is that some of its salient dynamical features are naturally accounted for starting from the Ising limit. The exactly soluble Ising limit itself has strongly-attractive magnon interactions based on simple domain-wall-counting. One does not expect such an Ising-based picture to be applicable to the \emph{low-energy} hydrodynamic Goldstone modes as $\lambda \to 1$, but our work shows that it does remain relevant for the \emph{high-energy} magnons.

In particular, the lowest non-trivial order in the Ising expansion already captures the physics of the roton mode---i.e. the dispersion along $|k_x|+|k_y|=\pi$---at a semi-quantitative level. Furthermore, we clarified its physical origin in terms of the properties of virtual bound states which mediate the magnon's hopping. On a more phenomenological level, the spectral function for the Heisenberg model obtained with DMRG shows that the anomalous local minimum at $\bm k = (\pi,0)$ grows monotonically when coming from the Ising limit. Moreover, even the strong continuum above this mode is continuously connected to spectral features near the Ising limit.

The spectral function on the geometry with $L_\textrm{circ}=10$ directly supports the point of view that the physics near $\bm k \approx (\pi,0)$ is beyond that of a perturbatively-dressed magnon: in Fig.~\ref{fig:isotropic_log} we saw that there is a continuum directly above the magnon which is \emph{not} a standard kinematic three-magnon continuum. Instead, the relevant physics is due to (quasi-)bound states arising from attractive interactions of magnons `sharing' their domain walls. This agrees with the insights of the CUT-based analysis\cite{Powalski15,Powalski17}. It would be interesting to further explore the potential link between the interactions arising in the CUT framework and that of the Ising-based picture. Remarkably, as far as the roton mode is concerned, a low-order Ising expansion gives predictions close to that of the CUT approach, but at this point it is not clear whether this is accidental or not.

Our study has also uncovered a curious spectral property. It turns out that magnons with $|k_x| + |k_y| = \pi$ are localized on their respective sublattices. This means that any operator localized on the A-sublattice cannot create a diagonal magnon associated with the B-sublattice. We have emphasized that this is rather unusual, since entanglement in the ground state would generically allow for it. Interestingly, such sublattice-localization is also predicted at low order in spin wave theory. It is not yet clear to what extent it is compatible with higher-order corrections in $1/S$.

Having established a link between the spectral properties of the Heisenberg model and an Ising-based picture, several questions can be raised. Firstly, does the latter simple picture also give an intuitive explanation for \emph{why} the (quasi-)bound states bunch up near $\bm k \approx (\pi,0)$? Secondly, as already mentioned in the introduction, the spectral features under discussion may not be uniquely described by one picture as opposed to another. Hence, we are not proposing the Ising-based picture as a complete framework, but rather as a very simple account of various features. Hence, it leaves open a link between the Ising limit and the other approaches that have been explored thus far. This could be interesting to investigate further. For example, a recent work interpreted the properties of the Heisenberg model in the context of the larger $J-Q$ model\cite{Shao17}, and it could be worthwhile to explore its interplay with an Ising anisotropy.

Finally, we note that although our analysis concerned zero temperature properties, much of the physics that we have discussed should stay relevant at low temperatures. In particular, while it is true that the Mermin-Wagner-Coleman theorem\cite{Mermin66,Coleman73} prevents ordering at $T\neq 0$, it has been calculated that magnons with momentum $|\bm k| \gg 1/\xi(T)$ (where $\xi(T) \sim \exp(\textrm{const.}/T)$) remain well-defined\cite{Dyson56,Harris71,Kopietz90,Ty90}. Efficiently extending the two-dimensional DMRG-based algorithm to finite temperatures remains a challenge for the future\cite{Vidal03,Verstraete04,White09,Barthel09,Stoudenmire10,Berta17,Hauschild2017}.

\section{Acknowledgements}

The authors would like to thank Ehud Altman, Sylvain Capponi, Fabian Essler, Efstratios Manousakis, Kai Schmidt and G\"otz Uhrig for helpful discussions. We are particularly indebted to Alexander Chernyshev for various discussions and useful comments on the manuscript, and to Jan Oitmaa for sharing the series expansion data of Ref.~\onlinecite{Zheng05} up to twelfth order. RV was supported by the German Research Foundation (DFG) through the Collaborative Research Centre SFB 1143 and FP acknowledges the support of the DFG Research Unit FOR 1807 through
grants no. PO 1370/2-1, TRR80, the Nanosystems Initiative Munich (NIM) by the German Excellence Initiative,  and the European Research Council (ERC) under the European Union's Horizon 2020 research and innovation program (grant agreement no. 771537).

\bibliography{SLAFHM_v2.bbl}

\appendix

\section{effective Hamiltonians to arbitrary order} \label{app:pert}

For completeness, we reproduce the general perturbative scheme that allows to obtain a well-defined effective Hamiltonian to any order. As in the main text, we consider a Hamiltonian of the form $H = H_0 + \lambda V$. Let $\mathcal H_0$ be the Hilbert space associated with the degenerate eigenvalue $E_0$ of $H_0$. Moreover, let $P_0$ be the projector onto $\mathcal H_0$, and $Q_0 = 1-P_0$. Note that we can decompose the total Hilbert space as $\mathcal H = \mathcal H_0 \oplus \mathcal H_0^\perp$.

Suppose that for $0 \leq \lambda \leq \lambda_c$, we can decompose the Hilbert space into $\mathcal H = \mathcal H_\lambda \oplus \mathcal H_\lambda^\perp$ in such a way that
\begin{enumerate}
	\item $\left(\mathcal H_{\lambda} \right)\large|_{\lambda=0} = \mathcal H_0$;
	\item $\mathcal H_\lambda$ is a smooth function of $\lambda \in [0,\lambda_c]$;
	\item the Hamiltonian respects the decomposition.
\end{enumerate}
Physically speaking, this formalizes the idea that we want the the energy levels of the sector we are interested to stay separated from the remaining levels; otherwise the idea of an effective Hamiltonian is misguided.

If those conditions hold, the work by Kato\cite{Kato49} and Takashi\cite{Takahashi77} showed that for the same range $0 \leq \lambda \leq \lambda_c$, one can explicitly construct a smooth unitary mapping $\Gamma_\lambda: \mathcal H_0 \to \mathcal H$ which maps the unperturbed eigenstates into the perturbed ones. Hence, the desired effective Hamiltonian on $\mathcal H_0$ is then simply $H_\textrm{eff} := \Gamma_\lambda^\dagger H \Gamma_\lambda$.

To perturbatively express $\Gamma_\lambda$ as a function of the known quantities $H_0$, $\lambda$ and $V_0$, it is useful to define a few other quantities. Firstly, let $P_\lambda :\mathcal H \to \mathcal H_\lambda$ be the projector onto $\mathcal H_\lambda$; we will derive a perturbative expression for this object as well. Secondly, define
\begin{align}
S^0 &:= - P_0, \\
S^k &:= \tilde G_0(E_0)^k := \left(Q_0 \frac{1}{E_0 - H_0} Q_0\right)^k  \quad (k \neq 0 ).
\end{align}
Note that $S^k$ is expressed in terms of \emph{known} quantities.

One can then derive\cite{Kato49,Takahashi77} that
\begin{equation}
P_\lambda P_0 = \sum_{n=0}^\infty \lambda^n  \sum_{\substack{k_1+k_2+\cdots+k_n = n,\\k_i \geq 0}} S^{k_1} V S^{k_2} V \cdots  S^{k_n} V P_0.
\end{equation}

Moreover, it can be shown that the following function then has the desired properties:
\begin{equation}
\Gamma_\lambda := P_\lambda P_0 \left( P_0 + \sum_{n=1}^\infty \frac{(2n-1)!!}{(2n)!!} [P_0 - P_0 P_\lambda P_0]^n \right).
\end{equation}
It can be proven that $\Gamma_\lambda$ as thus defined indeed satisfies $\Gamma_\lambda^\dagger \Gamma_\lambda = P_0$.

In terms of the above quantities, we thus have that
\begin{align}
H_\textrm{eff} &:= \Gamma_\lambda^\dagger H \Gamma_\lambda = E_0 P_0 + \lambda \; \Gamma_\lambda^\dagger V \Gamma_\lambda - \Gamma_\lambda^\dagger S^{-1} \Gamma_\lambda \; . \label{eq:appHeff}
\end{align}
The result in Eq.~\eqref{eq:appHeff}, namely that $\Gamma_\lambda^\dagger H_0 \Gamma_\lambda = E_0 P_0  - \Gamma_\lambda^\dagger S^{-1} \Gamma_\lambda$, is a direct consequence of $H_0 = H_0 P_0 + H_0 Q_0 = E_0 P_0 + (E_0Q_0 - S^{-1}) = E_0 - S^{-1}$ and the fact that $\Gamma_\lambda^\dagger \Gamma_\lambda = P_0$.

\section{finite-circumference effects for symmetry-equivalent points \label{app:analysis}}

\begin{figure}
	\includegraphics[width=0.7\linewidth]{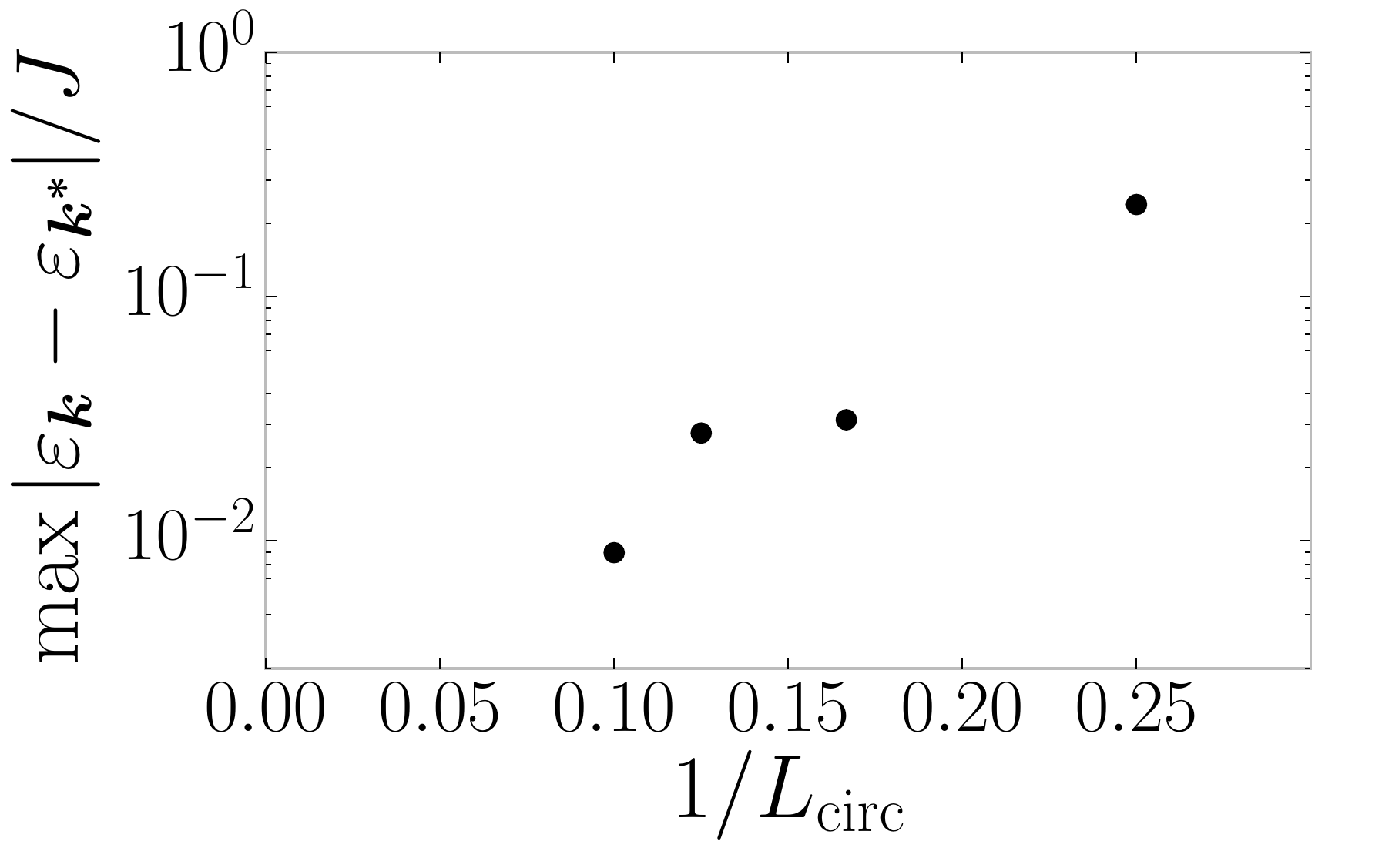}
	\caption{For a given $L_\textrm{circ}$, we consider the DMRG result for the one-magnon dispersion $\varepsilon_{\bm k}$ and find the maximum energy difference between momenta $\{\bm k, \bm {k^*}\}$ for which $\varepsilon_{\bm k} = \varepsilon_{\bm {k^*}}$ in the 2D limit. \label{fig:max_symmetry_inequiv}}
\end{figure}

The two-dimensional models enjoys symmetries which are broken when putting the model on a cylinder. Rotating the square lattice by $90^\circ$ gives an example. One can turn this curse into a blessing, since it gives us a direct probe of the finite-circumference effects. More precisely, suppose $\bm k$ and $\bm{k^*}$ are two distinct momenta which are symmetry-equivalent in the two-dimensional limit, but which are \emph{not} symmetry-equivalent on a cylinder with circumference $L_\textrm{circ}$. Hence, $|\varepsilon_{\bm k} - \varepsilon_{\bm{k^*}}|/J$ gives us a rough sense of how strong the finite-circumference effects are. Fig.~\ref{fig:max_symmetry_inequiv} plots the maximum of this over all pairs of symmetry-equivalent points. We see that it goes down as $L_\textrm{circ} \to \infty$, as expected.

\end{document}